\documentclass{aastex631}
\usepackage{CJK}

\usepackage{xcolor}
\usepackage{diagbox}
\usepackage{pifont}
\usepackage{lineno}
\begin{document}
	
\begin{CJK*}{UTF8}{gbsn}
		
\title{Locating heating channels of the solar corona in a plage region \\ with the aid of high-resolution 10830 \AA\ filtergrams}

\author{Parida Hashim}
\affiliation{Xinjiang Astronomical Observatory, CAS, Urumqi, 830011, China}
\author{Fangyu Xu}
\affiliation{Yunnan Astronomical Observatory, CAS, Kunming, 650216, China}
\author{Ya Wang}
\affiliation{Purple Mountain Observatory, CAS, Nanjing, 210023, China}
\author{Weijie Meng}
\affiliation{Yunnan Astronomical Observatory, CAS, Kunming, 650216, China}
\affiliation{University of CAS, Beijing, 100049, China}
\author{Jinhua Shen}
\affiliation{Xinjiang Astronomical Observatory, CAS, Urumqi, 830011, China}
\author{Yingna Su}
\affiliation{Purple Mountain Observatory, CAS, Nanjing, 210023, China}
\author{Jianping Li}
\affiliation{Purple Mountain Observatory, CAS, Nanjing, 210023, China}
\author{Zhenyu Jin}
\affiliation{Yunnan Astronomical Observatory, CAS, Kunming, 650216, China}
\author{Haisheng Ji}
\affiliation{Purple Mountain Observatory, CAS, Nanjing, 210023, China}
\correspondingauthor{Haisheng Ji, Zhenyu Jin}		
\email{jihs@pmo.ac.cn, kim@ynao.ac.cn}
\begin{abstract}
In this paper, with a set of high-resolution He I 10830 \AA\ filtergrams, we select an area in a plage, very likely an EUV moss area, as an interface layer to follow the clues of coronal heating channels down to the photosphere. The filtergrams are obtained from the 1-meter aperture New Vacuum Solar Telescope (NVST).  We make a distinction between the darker and the brighter regions in the selected area and name the two regions enhanced absorption patches (EAPs) and low absorption patches (LAPs). With well-aligned, nearly simultaneous data from multiple channels of the AIA and the continuum of the HMI on board SDO, we compare the EUV/UV emissions, emission measure, mean temperature, and continuum intensity in the two kinds of regions.  The following progress is made: 1) The mean EUV  emissions over EAPs are mostly stronger than the corresponding emissions over LAPs except for the emission at 335 \AA.  The UV emissions at 1600 and 1700 \AA\ fail to capture the difference between the two regions. 2) In the logarithmic temperature range of 5.6-6.2, EAPs have higher EUV emission measure than LAPs, but they have lower mean coronal temperature. 3) The mean continuum intensity over EAPs is lower.  Based on the above progress, we suggest that the energy for coronal heating in the moss region can be traced down to some areas in intergranular lanes with enhanced density of both cool and hot material. The lower temperature over the EAPs is due to the greater fraction of cool material over there.
\end{abstract}

%------------------------------------------------------------------------------------------------------------------------------------------------
\section{Introduction} \label{sec:intro}
%2020A&A...644A.130C(Chitta), 2010A&A...521A..21O(O'Dwyer), 2019ApJ...870..114S(Sim{\~o}es})
How the solar or stellar corona is heated to millions of degrees is an unsolved problem in the field of astrophysics \citep{Klimchuk2006SoPh, Schrijver2000ssma.book, 2005psci.book.....A}. Most researchers agree that the corona is probably heated in two primary ways: dissipation of MHD waves and nano-flares \citep{Parker1988ApJ} that occur directly in the corona. The latter was believed to be caused by magnetic reconnection in the braided coronal magnetic field. Nanoflares are often associated with impulsive heating \citep{Warren2003ApJ, Patsourakos2006ApJ, Tripathi2010A&A}, while wave heating is believed to be in a steady state \citep{Reale2000ApJ, 2003ApJ...590..547A, Brooks2009ApJ}. Both kinds of heating mechanisms have their support from contemporary observations, e.g., even with the highest observations from Hi-C in the extreme-ultraviolet (EUV), the opinions regarding steady-state heating or impulsive heating are co-existing \citep{2013ApJ...770L...1T, Graham2019ApJ, Warren2020ApJ}. \cite{Aschwanden2007ApJ} pointed out that energy for heating the corona certainly comes from the lower atmosphere, which includes the lower transition region (TR), the chromosphere, and the photosphere. Also, more and more high-resolution observations from the lower atmosphere reveal that the heating process is at small spatial and temporal scales via hot expulsions, spicules, or waves generated directly from the photosphere, leaving responses in the TR and chromosphere \citep{Tian2014ApJ, Peter2014Sci, 2019Sci...366..890S, Ji2012ApJ, Ji2021RAA, Hashim2021RAA}. A key issue before resolving the problem lies in determining where the heating channels are located.  \\

%The observational facts raise another key issue of how to reconcile the dynamic lower atmosphere and the belief in the steadiness of coronal heating that is reinforced by recent high-resolution EUV observations from EUI on board Solar Orbiter \citep{2023ApJ...957...25J}. \\ Ishikawa et al. {2021SciA....7.8406I} found that large areas of sunspot plages are magnetically unipolar as measured in the cores of Mg II resonance lines, formed in the lower TR under low plasma-β conditions. 

The study of TR emission provides a direct way to investigate coronal heating sources due to the fact that many TR lines are very sensitive to heating events, and their emissions are sometimes confined to a thin atmospheric layer.  For these reasons,  most papers on coronal heating focus particularly on the emissions from the so-called EUV moss regions. An EUV moss region appears as a bright, reticulated pattern with dark inclusions above a solar active region plage \citep{1999ApJ...519L..97B} and was first seen in images from the Normal Incidence X-ray Telescope \citep{Peres1994ApJ}. Observations using TRACE and soft X-ray telescopes indicate that an EUV moss region corresponds to the thin area at the footpoints of hot coronal loops with multi-million-degree plasma \citep{1999SoPh..190..409B, 2000ApJ...537..471M, 2003ApJ...590..547A, 2013ApJ...770L...1T}. Due to its connection with high-temperature coronal loops, which disappear in EUV images, it was originally thought of as an ideal place for studying the heating of TR via conductive heating from above.  \cite{Fletcher1999AAS} determined the plasma parameters of a moss region, including its electron density, and obtained its differential emission measure (DEM).  The DEM analysis confirmed that the moss region is composed of a million-degree plasma, with temperatures ranging from $ 10^{5}$K to $ 10^{6}$ K.  In addition, solar plage regions are usually associated with unipolar magnetic fields, challenging the heating models of magnetic reconnection in the lower atmosphere \cite[e.g.][]{2024ApJ...960..129J}. For a moss region, there may exist a classical transition region with the energy flow being dominant by downward heat conduction. For example, there is a good association between EUV emission from the TR and  H$\alpha$ and Ly$\alpha$ emission from the chromosphere \citep{2003ApJ...595L..63D, 2001ApJ...563..374V}. However, Ca II K bright kernels fail to be co-spatial with the EUV emissions. Also, \cite{2001ApJ...563..374V} extensively discusses the fact that the Ly$\alpha$ emission ``below'' the EUV 171 \AA\ emission is much too bright (by a factor of 500)  and indicates that the excess brightness of the chromosphere can probably not be explained by classical thermal conduction from the upper TR and corona \citep[see also][]{2008ApJ...683L..87J, 2008ApJ...687.1388J}.  Therefore, coordinated observation of the chromosphere with high resolution is important for understanding the heating of the chromosphere and its coupling with the TR, which is usually carried out in the wavelengths of H$\alpha$ and Ca II K lines on the ground. \\

On the other hand, high-resolution imaging at He I 10830 \AA\ has proven to be a powerful tool for locating heating sources since helium excitation is only sensitive to high-energy photons with a wavelength less than 504 \AA\ or high-energy electrons and ions whenever the temperature is above 20,000 K. The two excitation mechanisms respectively correspond to photo-ionization followed by recombination \citep[the ``PR'' mechanism][]{1975ApJ...199L..63Z} or direct collision \citep{1975MNRAS.170..429J, 2000ApJ...535..438A, 2004ApJ...606.1239P, 2005A&A...432..699D}. It is accepted that both kinds of mechanisms work for helium excitation, and they have obtained their support from observation \citep[e.g.:][]{2016A&A...594A.104L, 2016ApJ...833..250W, 2014ApJ...793...87Z, 2004ApJ...606.1258J}.   Over the solar disk, the morphology of 10830 \AA\ absorption in the high-resolution images is quite inhomogeneous, showing structures of various scales from large-scale coronal holes to small-scale darkened fibrils or jets, including micro-jets \citep{Hong2017RAA, Hashim2021RAA, 2022ApJ...928..153H}. The spatial structure in 10830 \AA\ reflects the change in opacity, which results in the darker and brighter regions interspersed in a 10830 \AA\ image. The brighter regions are those places where the photospheric continuum shines through. In a 3D radiation-MHD simulation, \cite{2016A&A...594A.104L} show that the chromospheric electron density causes the change in opacity.  However, their simulations are not representative of strongly magnetically active areas like plage regions.   \\ 

The dual features of the darker regions and the brighter regions make them ideal for constructing a connection between the photosphere and the TR. The special characteristic has enabled several investigations to pin down somewhere in solar intergranular lanes as the heating source from jet-like dynamical events by magnetic reconnection or other unknown mechanisms \citep[e.g., ][]{2022ApJ...928..153H}. Recent studies by \cite{Hashim2021RAA, Ji2021RAA} for an EUV moss region demonstrate the spatiotemporal correlation among the quasi-periodic enhancement of  He I 10830\AA\ absorption in intergranular lane areas, EUV brightenings, and magneto-acoustic oscillations in a sausage mode. With the aid of comprehensive EUV emissions, a detailed analysis of the thermal structures around the darker area may shed light on the problem of coronal heating as well as the excitation mechanism of orthohelium atoms of these regions. \\

In this paper, by using high-resolution 10830 \AA\ narrow-band filtergrams from NVST, we give a detailed analysis for a plage area, which is very likely an EUV moss region, focusing on the difference of EUV emission or emission measure from the TR, the mean temperature, and the continuum emission from the photosphere over the darker and the brighter places on the 10803 \AA\ filtergrams. For completeness, we also include UV emissions at 1600 and 1700 \AA\ in the analysis. The observation and data analysis are given in Section \ref{sec: OBSERVATION AND DATA ANALYSIS}, and the results are presented in Section \ref{sec: RESULTS}. Then, our discussions and conclusions are presented in Sections \ref{sec: Discussion} and \ref{sec: Conclusion}. \\
%%%%%%%%%%%%%%%%%%%%%%%%%%%%%%%%%%%%%%%%%%%%%%%%%%%%%%%%%%%%%%%%%%%%%%%%%%%%

%%%%%%%%%%%%%%%%%%%%%%%%%%%%%%%%%%%%%%%%%%%%%%%%%%%%%%%%%%%%%%%%%%%%%%%%

\section{OBSERVATION AND DATA ANALYSIS} \label{sec: OBSERVATION AND DATA ANALYSIS}

The 10830 \AA\ filtergrams analyzed in this paper were obtained from the 1-meter aperture New Vacuum Solar Telescope \citep[NVST: ][]{Liu2014RAA, Yan2020ScChE} at the Fuxian Solar Observatory (FSO) on 2022 September 27 from 06:14:30 UT to 07:05:25 UT. The narrow-band 10830 \AA\ filter (bandpass 0.5 \AA), manufactured at Nanjing Institute of Astronomical and Optical Technology, is the same as the one borrowed to use at Big Bear Solar Observatory (BBSO) \citep{Ji2012ApJ}.  The near-infrared detector was assembled at the Yunnan Astronomical Observatory, and the Kunming Institute of Physics developed the near-infrared detector chip. The detector has a $640\times512$  FPA format with 15 $\mu$m square pixels and a well depth of approximately 20500 $e^{-}$, digitized using a 16-bit analog-to-digital converter with a gain of $\sim$ 0.28 $e^{-}$ per count. All counts have been normalized after dark current and flat field processing. The 10830 \AA\ filtergrams analyzed in this paper are from the first light of the detector. The high-resolution statistical reconstruction of the speckle image was carried out with the Non-rigid Alignment Based Solar Image Reconstruction (NASIR) \citep{2022RAA....22i5005L} method, and a high-resolution image was obtained to the diffraction limit of the NVST. \\

We also used simultaneous space observations data from Atmospheric Imaging Assembly \citep[AIA: ][]{2012SoPh..275...17L} and Helioseismic and Magnetic Imager \citep[HMI: ][]{2012SoPh..275..229S} on board Solar Dynamics Observatory \citep[SDO: ][]{2012SoPh..275....3P}. The extreme ultraviolet (EUV) and ultraviolet (UV) images were observed by the AIA,  having a full image size of $4096 \times 4096$ pixels, with a pixel size of  $0.6'' $. The seven EUV bandpasses of AIA, 94 \AA\ (Fe XVIII line), 131 \AA\ (Fe VIII and Fe XXI lines), 171 \AA\ (Fe IX line), 193 \AA\ (Fe XII and Fe XXIV lines), 211 \AA\ (Fe XIV line), 304 \AA\ (He II line), 335 \AA\ (Fe XXVI line), have different responses to different temperatures \citep{2010A&A...521A..21O}. Its two UV bandpasses, 1600 and 1700 \AA, sample emissions from the photosphere and chromosphere up to the lower transition region depending on different atmospheric conditions \citep{2019ApJ...870..114S, 2020A&A...644A.130C}. The spatial resolution for all bands is $1.2'' $ and with a time cadence of 12 s. SDO/HMI is the first space-based instrument to map the full-disk photospheric vector magnetic field with high cadence and continuity. The spatial and temporal resolutions of HMI observations are  $1.5'' $  and 45s, respectively. \\

To align NVST 10830\AA\ images with HMI continuum images, we used sunspots, especially bright granulations, to achieve a precise co-alignment. Since the brighter regions of the helium 10830\AA\ images have rich granulation information, the accuracy of image alignment between the chromosphere and photosphere can be better than $0.2''$. 
%%%%%%%%%%%%%%%%%%%%%%%%%%%%%%%%%%%%%%%%%%%%%%%%%%%%%%%%%%%%%%%%%%%%%%%
\section{RESULTS}\label{sec: RESULTS}

\begin{figure}
\plotone{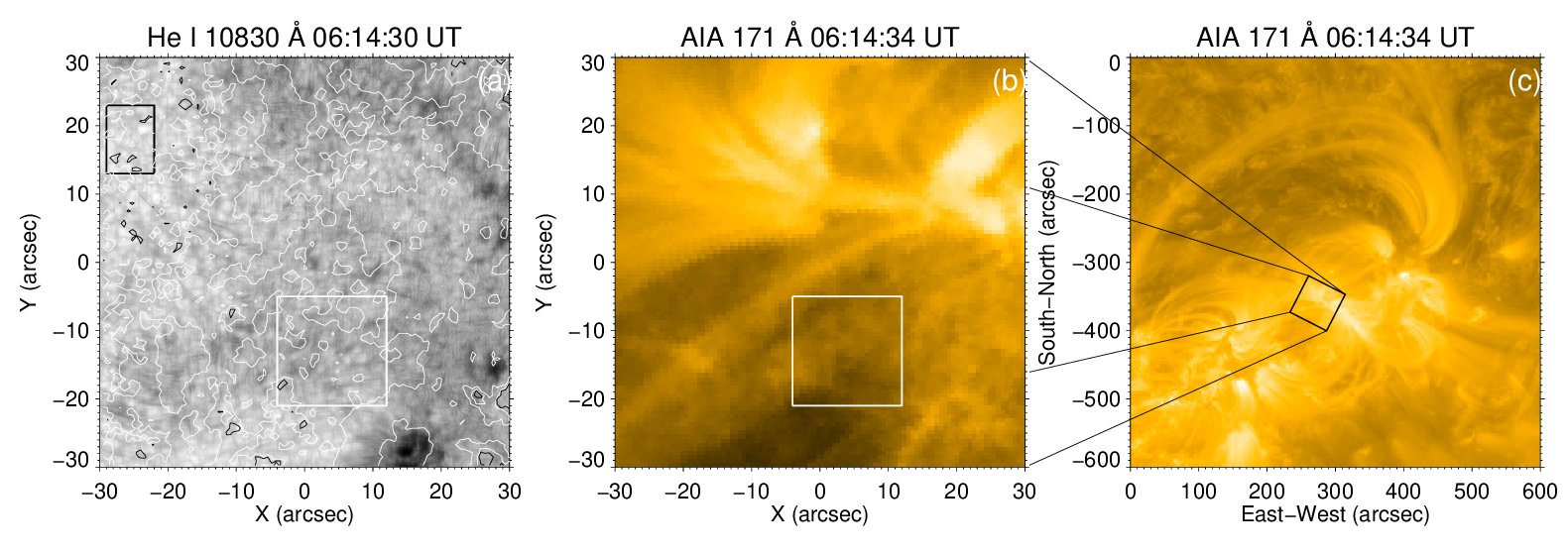}
\caption{An overall view of the observations for the solar active region NOAA 13105. Panels (a-b) give a sample high-resolution He I 10830 \AA\ narrow-band (bandpass 0.5 \AA) image observed by NVST  and an EUV image observed by the SDO/AIA at Fe IX 171 \AA\ with the same field of view (FOV).  The surrounding context of the observation is shown in panel (c) with the larger FOV. The white boxes outline the region of interest (ROI) being analyzed in this paper, which is also the FOV for Figs. 2b, 3, and 5. The black box in panel a indicates the background with the least 10830 \AA\ absorption, and the light curve for the background is given in Figure 2a (the gray one). Overlaid contours in panel (a) represent the line of sight (LOS) magnetic field with values of 10, 100, 1000, and -10 G, respectively. The contours for -10 Gausses are specially given in black.}
\label{fig:fig1}
\end{figure}

On September 27, 2022, an observation was made targeting the part of the active region NOAA 13105 from 06:14:30 to 07:05:25 UT.  The observation produced 262 images, giving the frame cadence of about 12 seconds. Figures 1a-b give a sample He I 10830 \AA\ narrow band image and a corresponding Fe IX 171 \AA\ image with the same field of view ($60'' \times 60''$). The 10830 \AA\ image covers a surrounding plage region of the active region, including a sunspot. We overlaid the line of sight (LOS) magnetic field observed by HMI in Figure 1a. It can be found that the areas of plages and the sunspot are predominantly unipolar with a positive magnetic field. Figure 1c gives an intermediate zoom level of the AIA image, giving its position and orientation on the solar disk and its surrounding information.  Compared with the observations made with GST \citep{Ji2012ApJ}, the 10830 \AA\ image obtained by NVST is quite similar, though with degraded spatial resolution. Lots of prominent absorbing features from the upper chromosphere and background of the photosphere with sunspots and granules co-exist on the 10830 \AA\ image. The photospheric features give us perfect references for the co-alignment with the continuum images of HMI/SDO.  From the EUV images, we see that most plage areas are projected with overlying transition region loops.  To use the 10830 \AA\ absorption features as a connection between the photosphere and the emission nature of the transition region emission, we choose the white-boxed area in panels a and b as the region of interest (ROI) with the FOV of $16'' \times 16''$ for a detailed analysis. The selected area has avoided overlying EUV loops and the nearby sunspot, and we see that, in the selected area, EUV emission is low-lying and has a spongy appearance, merits the criteria for an EUV moss region (also see Fig. 3).  As the observation in the previous study, the movie made of the 10830 \AA\ images shows many quasi-periodic small-scale mass ejections. The quasi-periodic small-scale mass ejections were found to be associated with magneto-acoustic oscillations \citep{Hashim2021RAA, Ji2021RAA}. The relationship between the small-scale mass ejections and magneto-acoustic oscillations could be investigated in a separate research. \\

\begin{figure}
\plotone{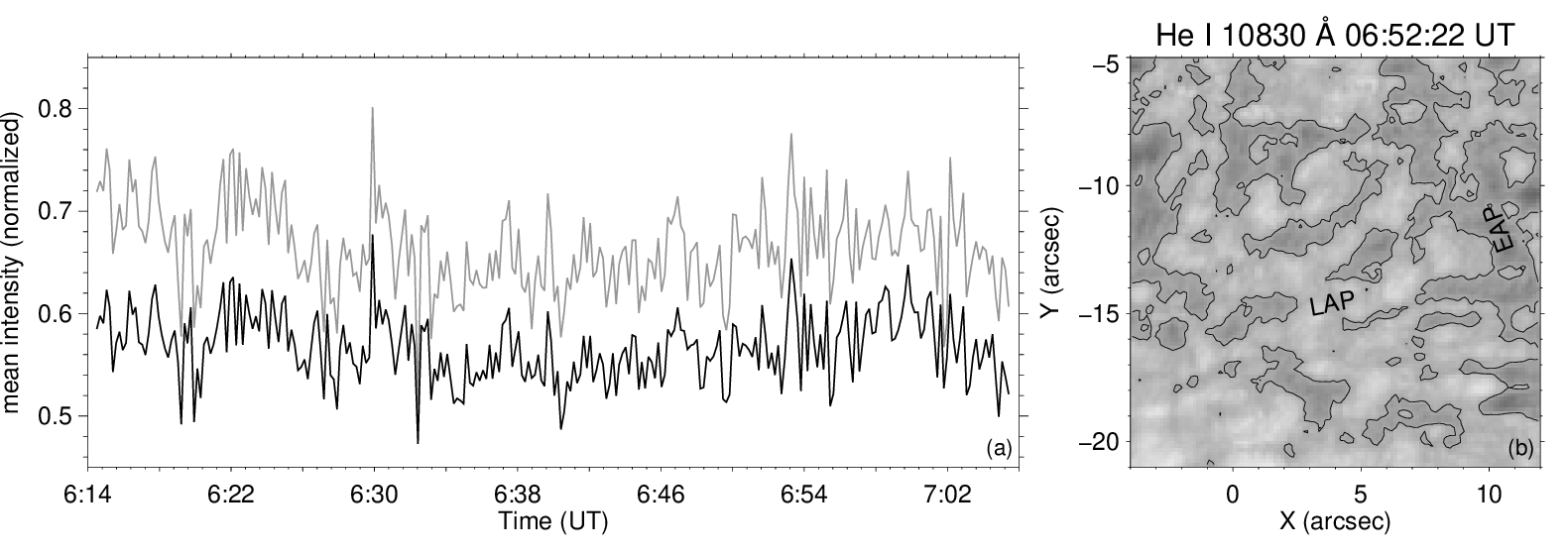}
\caption{The left panel gives two kinds of time profiles obtained from the 10830 \AA\ images. The gray (upper) one is for the mean intensities inside the black box in Figure 1a with the least 10830 \AA\ absorption, while the black (lower) one is for the mean intensities inside the ROI (the white boxes in Figure 1a-b). The right panel gives a sample He I 10830 \AA\ image in the ROI overlaid with a contour level to pick out the darker regions and name them enhanced absorption patches (EAPs). Other regions (brighter regions) are named low absorption patches (LAPs). The contour level for distinguishing EAPs and LAPs in each image is given by 0.9 of the average bandpass intensity of the ROI. To show the spatiotemporal change of EAPs and LAPs, an online animation is given.} 
\label{fig:fig2}
\end{figure}

To follow the 10830 \AA\ absorption features, we divide 10830 \AA\ images inside the ROI into two kinds of regions: enhanced absorption patches (EAPs) and low absorption patches (LAPs). They are viewed as the darker and the brighter patches interspersed together in the ROI. Put simply, EAPs are the darker regions, and LAPs are the brighter regions in the filtergrams, reflecting the spatial inhomogeneity of 10830 \AA\ opacity.  The classification has to be made quantitatively using the brightness, but the brightness observed on the ground is affected by the variations of the transmission through the Earth's atmosphere. To show how the variations affect the evaluation of the 10830 \AA\ absorption, in Figure 2a, we give two kinds of time profiles obtained from the 10830 \AA\ images. The gray (upper) one is for the mean intensities in the area with the least 10830 \AA\ absorption (the black box in Figure 1a, taken as a background ), while the black (lower) is for the mean intensities inside the ROI (the white box in Figure 1a-b).  We see that they are vertically shifted but highly correlated. To pick out EAPs at a certain time, we follow the method of \cite{Hong2017RAA}, which defines EAPs as those darkened areas within the contours of  $\sim$ 90\% average bandpass intensity at the time. In the paper of \cite{Hong2017RAA}, the contour value is equivalent to the average intensity along thin dark loops connecting two plage regions of the active region. The right panel of Figure 2 shows the distribution of EAPs in the ROI at 06:52:52 UT, which are inside the defined contour levels for absorption of the time.  The method gives changing criteria for EAPs for different images, but it has the merit of overcoming intensity fluctuation of ground-based observation caused by the atmosphere.  In this way, the summed area of all EAPs occupies approximately one-third of the total area of the ROI all the time,  though the contour levels and locations of EAPs keep changing from image to image.  To show the constant spatial change of EAPs and LAPs, an online animation is made for Fig. 2. From the animation, we see that EAPs keep changing to LAPs and vice-versa. The stable area fraction results in a stable sample size, which will have a minimum impact on the comparison study regarding average EUV emission, emission measure, and temperature of the two different regions. In the following, we compare these quantities for the two kinds of regions, exploring the role of EAPs in helping us understand the heating of the upper atmosphere.   \\

\begin{figure}
	\plotone{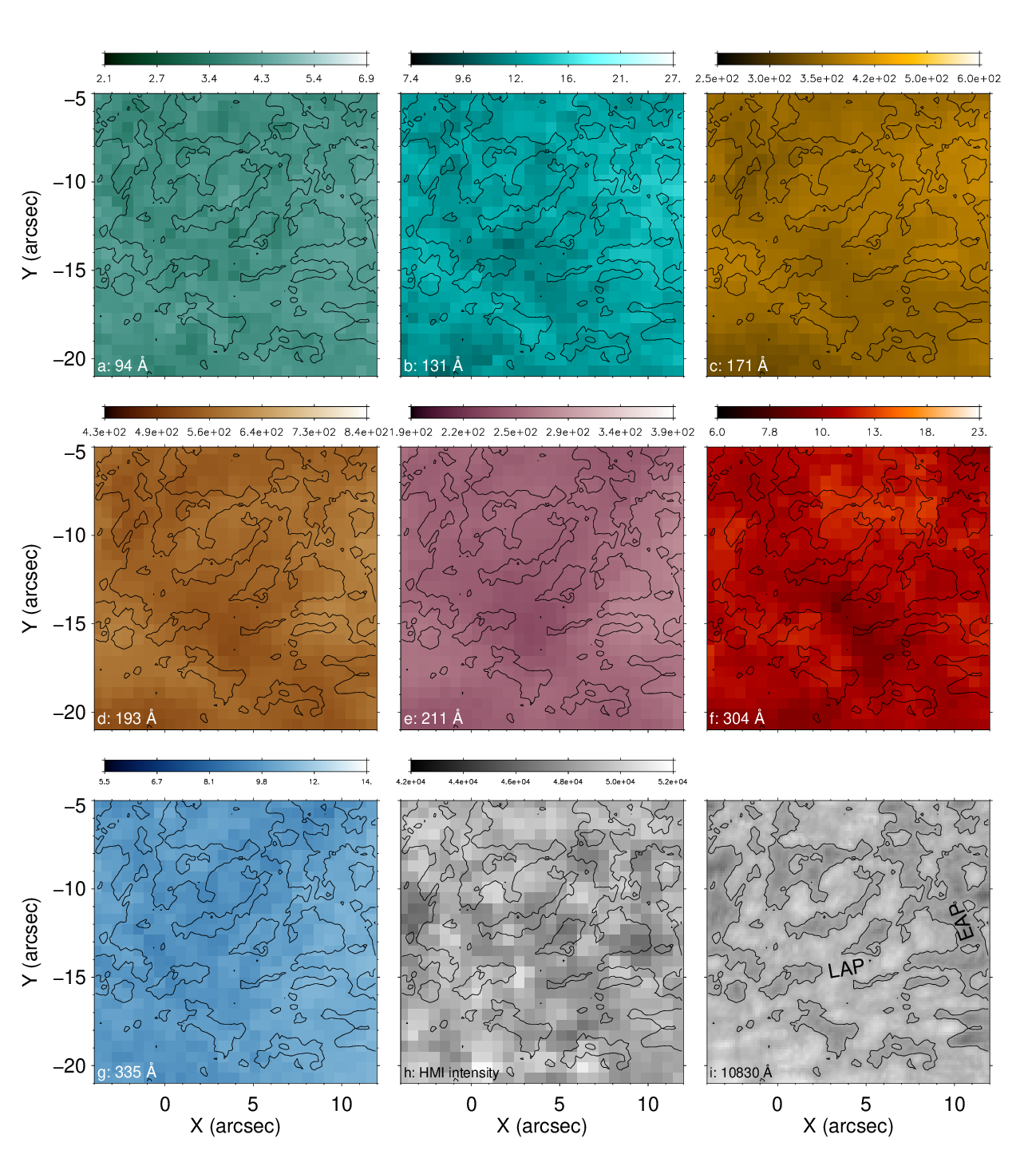}
	\caption{Panels a-h give snapshots of AIA images of different wavelengths and an HMI continuum intensity image in the ROI. All the EUV images, including the continuum image, were taken at around 06:56:20 UT, nearly simultaneous with the 10830 \AA\ image in Fig. 2b. For readers' convenience, here we redraw Fig. 2b in panel (i).  Overlaid on all the images are the contour level to define the 10830 \AA\ EAPs of the moment, the same contour level as in Figure 2b. The displaying values for the EUV and the continuum images are given with color bars above.} 
	\label{fig:fig3}
\end{figure}

To investigate the spatial correlation of EUV emission and 10830 \AA\ absorption in the ROI, Figure 3(a)-(g) gives 7 images as observed in different EUV wavelengths in the ROI. From the EUV images, we see that the EUV emission has the appearance of a moss region.  To see the spatial correlation of the EAPs with EUV emissions, the EUV images are overlaid with contours of the EAPs from nearly simultaneous 10830 \AA\ filtergram that is also put in Fig. 7i for the purpose of an immediate comparison.  We can see that, for 131, 171, and 304 \AA, the EUV emission inside the areas outlined by contours is brighter than that of areas out of the contours. For these EUV wavelengths, almost all EUV bright areas are within the contours of the 10830 \AA\ EAPs, and the contours stay away from regions of weak EUV emission. Here, it is worth noting that the EAPs do not correspond to dark inclusions of the moss region. The nice spatial correlation becomes a little bit poor for 94, 193, and 211 \AA\ images, and it vanishes for 335 \AA\ images. Figure 3(h) gives an HMI continuum intensity map overlaid with the same contours. In order to distinguish granules from intergranular lanes, the image's contrast has been especially increased. However, due to insufficient spatial resolution, here it's still hard to draw a conclusion about the spatial relationship between the EAPs and intergranular lanes like the one made by Hong et al. (2017), who claimed that the EAPs usually lie on the intergranular lanes. \\

\begin{figure}
	\plotone{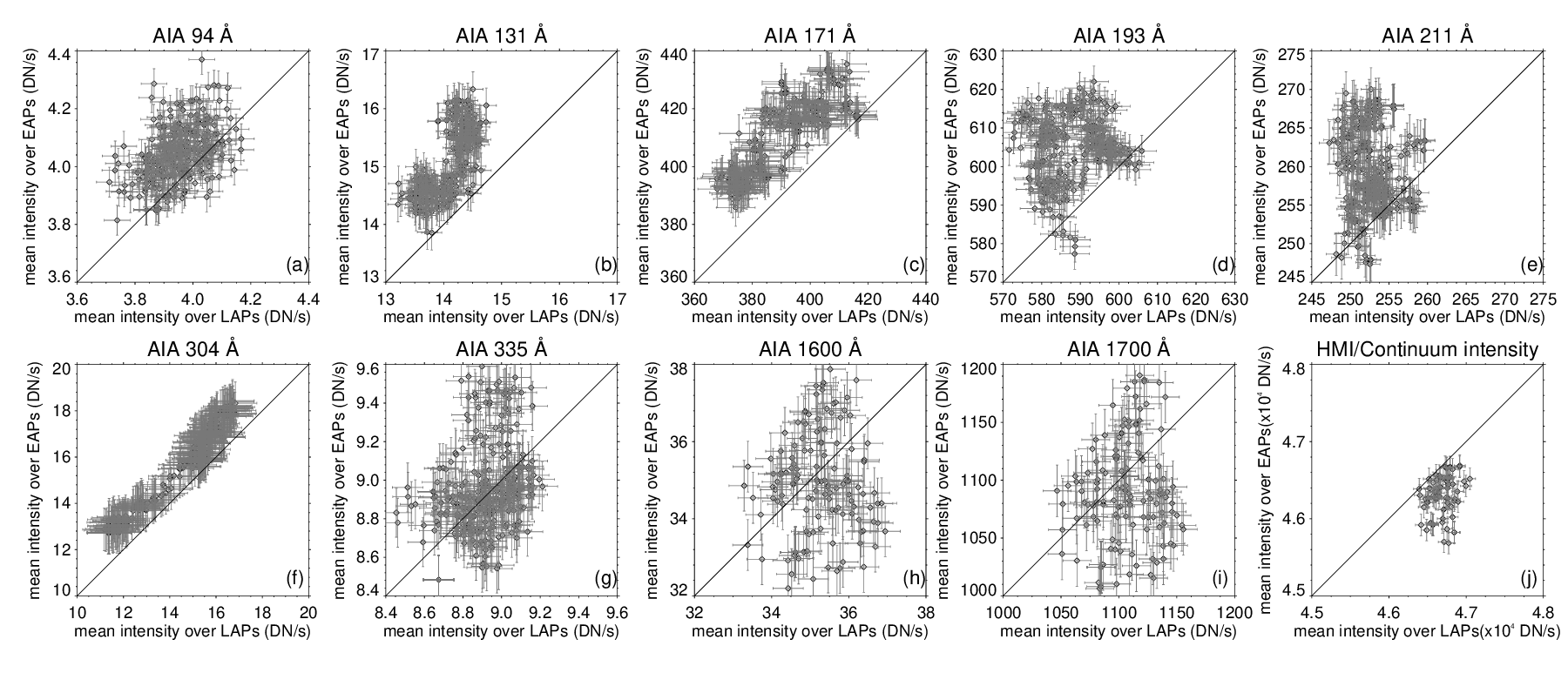}
	\caption{Scatter plots of the mean intensity of EUV/UV emission (panels a-i) and continuum intensity (panel h) over the EAPs versus the corresponding mean intensity over the LAPs. In this figure and Figs. 6 and 7, the lengths of the over-plotted error bars equal the standard deviation obtained from their time profiles.} \label{fig:fig4}
\end{figure}

To further investigate the spatial correlation of the EUV/UV and continuum emission and 10830 \AA\ absorption in the ROI during the whole observing time period, we present a series of scatter plots of their mean intensity over the EAPs versus the corresponding one over the LAPs. The plots are shown in Figure 4, consisting of all points from corresponding frames. In this figure and Figs. 6 and 7, the lengths of the over-plotted error bars equal the standard deviation obtained from the time profiles of the mean EUV/UV and continuum intensity over the EAPs or LAPs.  Figure 4 indicates an agreement with the results in Figure 3.  We can find that, nearly at all times, EUV emissions at 131, 171,  and 304 \AA\ over the EAPs are more intense than those of other regions (Fig. 4b-c and f). For 94, 193, and 211 \AA\, the conclusion still holds true most of the time (Fig. 4a and d-e). But for line 335 \AA, such a trend vanishes (Fig. 4g). There is no difference between the two regions' 1600 and 1700 \AA\ emissions, either (Fig. 4h-i). Figure 4h shows a similar scatter plot comparing the two kinds of HMI continuum intensity (representing the brightness of the photosphere). We see the difference between the photosphere's brightness over the EAPs and LAPs.  Over the EAPs, the photosphere's brightness is slightly darker than over the LAPs. The result actually supports the picture that the EAPs are usually above intergranular lanes (Hong et al. 2017).  \\

\begin{figure}
	\plotone{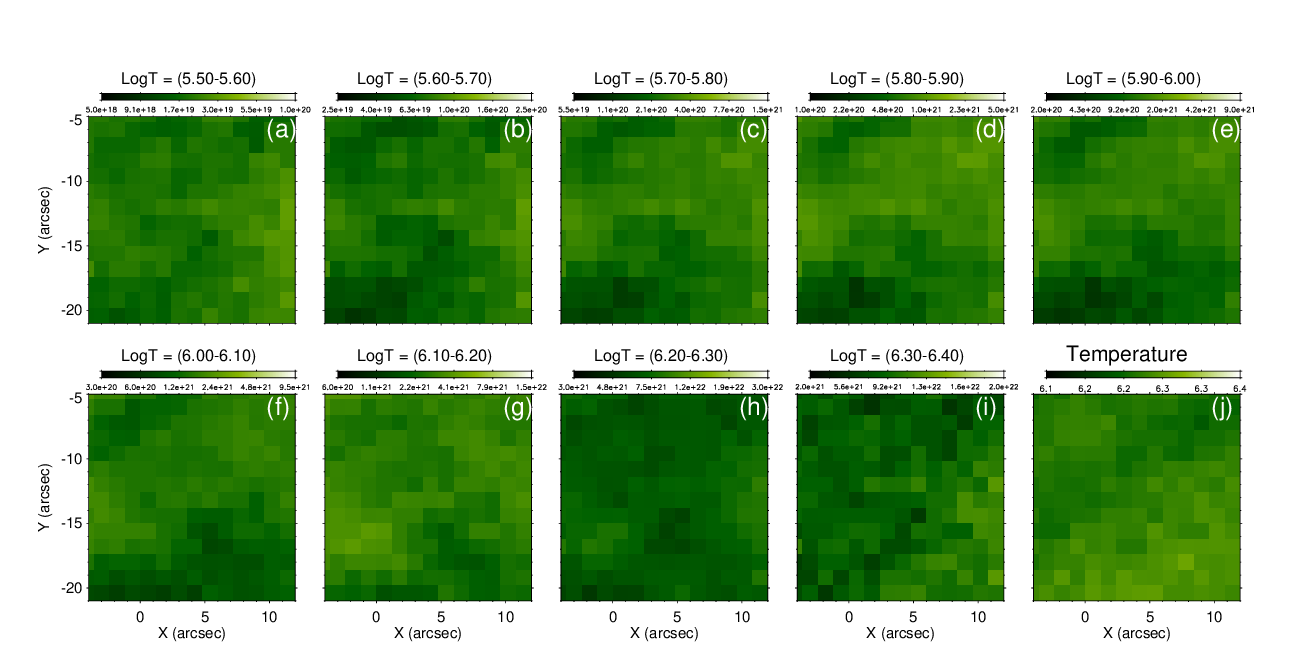}
	\caption{Emission measure maps at binned logarithmic temperature ranges (panels a-i) and a temperature map
(panel j). The displaying values for each panel are given in a color bar above.} \label{fig:fig3}
\end{figure}

To inspect the spatiotemporal thermal behavior of the moss region, we carried out an analysis of differential emission measure \citep[DEM:][]{2015ApJ...807..143C, 2018ApJ...856L..17S} to the region using the six AIA passbands.  The starting logarithmic temperature range is set to be [5.5, 5.6], and all following ranges have an ongoing 0.1 increment until ending at [6.3, 6.4], as illustrated in panels a-j of Figure 5. Temperature distribution derived from the DEM analysis is given in panel j of Figure 5, and the region's average temperature is about 1.78 MK. \\

\begin{figure}
	\plotone{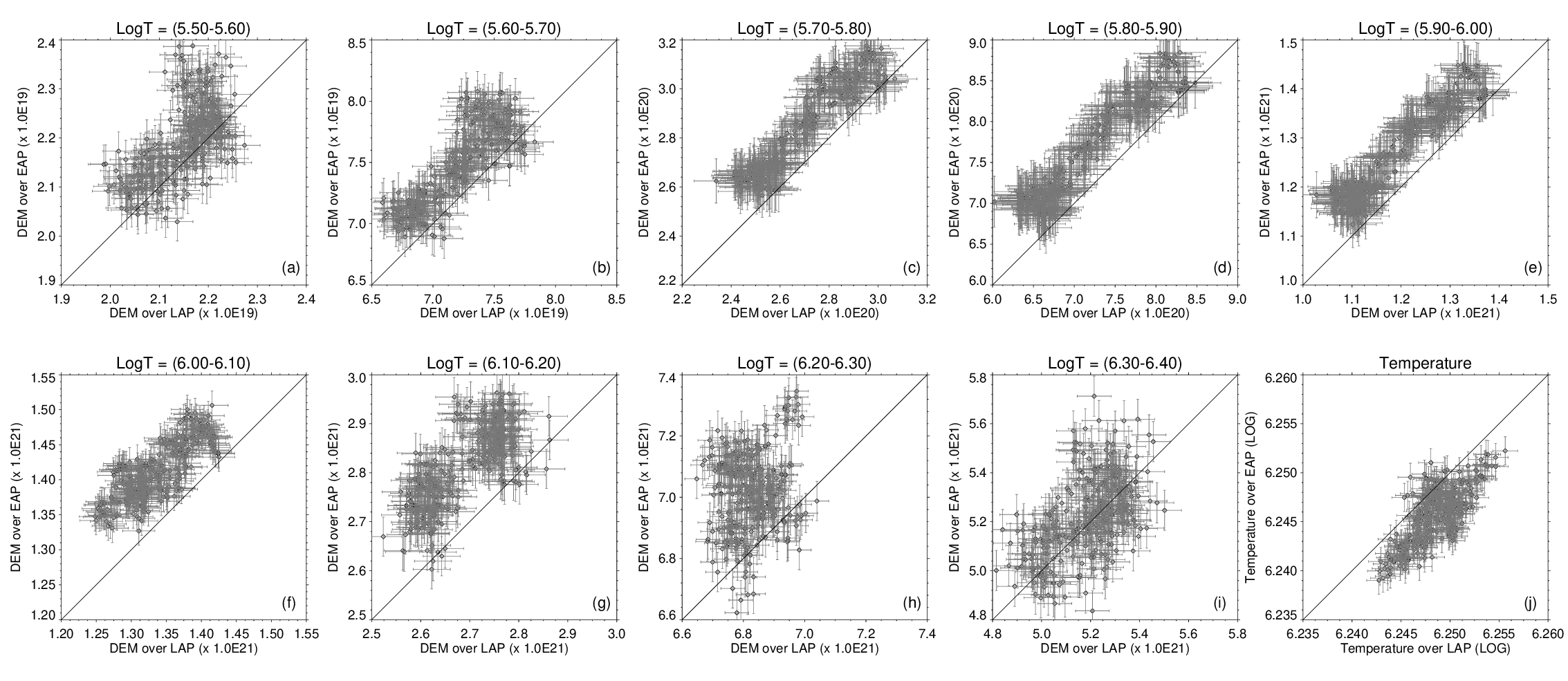}
	\caption{Scatter plots of the mean EUV emission measure at different temperature ranges (panels a-i) and mean temperature (panel j) over the EAPs versus the corresponding values over the LAPs.} \label{fig:fig3}
\end{figure}

Similar to Figure 4, Figure 6a-j gives a series of scatter plots of the mean EUV emission measure and mean temperature over the EAPs versus corresponding values over the LAPs. We can see the mean emission measure in the logarithmic temperature range of 5.6 to 6.2 is mostly larger over EAPs. On the contrary, the mean temperature over EAPs is mostly lower. We then use the temperature to divide the ROI into relatively high-temperature regions and low-temperature regions with equal areas. We inspect if there is a difference between the two regions by comparing the mean 10830 \AA\ intensity and mean emission measure.  The two scatter plots in Figure 7a-b present the results. It can be seen that the mean 10830 \AA\ intensity over the high-temperature regions is mostly intenser than over the low-temperature regions (Figure 7a). The result and Fig. 6j support each other from two aspects, showing that the EAPs are spatially associated with lower-temperature regions. In the same way, Figure 7b shows that the mean EUV emission measure over the low-temperature regions is slightly higher than over the high-temperature regions. It can be seen that, over the lower-temperature regions, the mass density of both cool and hot material is higher. 

\begin{figure}
	\plotone{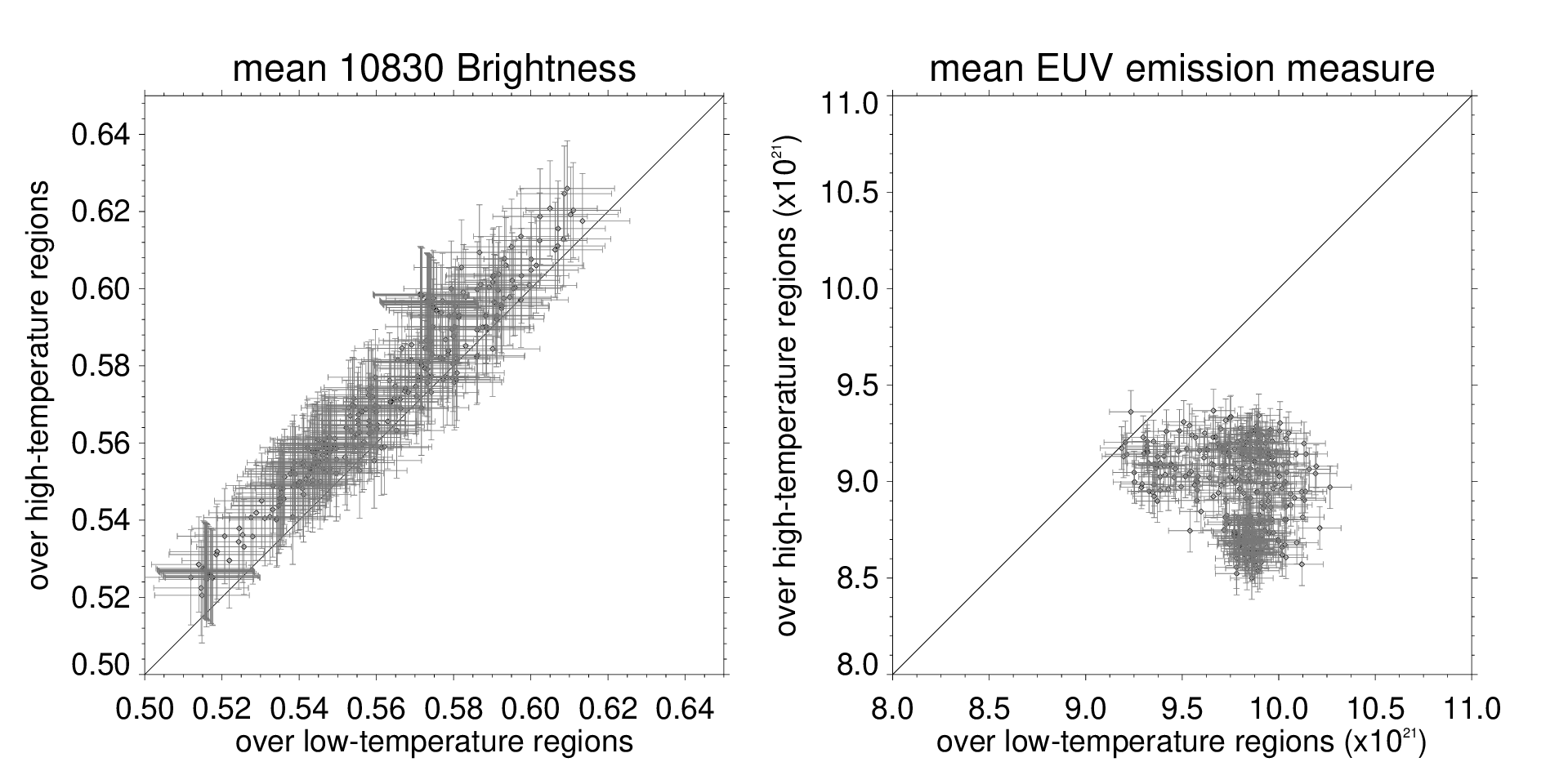}
	\caption{Comparison for the mean 10830 brightness and EUV emission measure over the high-temperature and low-temperature regions. The EUV emission measure is totaled in the logarithmic range of 5.6-6.2.}
 \label{fig:fig7}
\end{figure}
%%%%%%%%%%%%%%%%%%%%%%%%%%%%%%%%%%%%%%%%%%%%%%%%%%%%%%%%%%%%%%%%%%%%%%%%%%
\section{Discussion} \label{sec: Discussion}

High-resolution He I 10830 \AA\ filtergrams are quite inhomogeneous with the darker and the brighter regions. The darker regions represent the more 10830 \AA\ absorption in these areas. Since helium excitation is only sensitive to high-energy photons or high-energy particles, they provide us with an interface layer for studying the coupling between the lower atmosphere and the transition region.  For the moss area analyzed in this paper, we name the darker and the brighter regions enhanced absorption patches (EAPs) and low absorption patches (LAPs).  The brightness criterion for defining the EAPs at a given time is set to include those pixels whose brightness is less than 90\% of the average for the area. The 90\% is actually empirical. By making the criterion connect to the average brightness of the ROI at each moment, we overcome the effects caused by the fluctuation of transmission of the earth's atmosphere.  In this way, EAPs steadily occupy about 1/3 of the total area of the moss area with steady sample pixel numbers for statistics. Using multiple channel observations of the AIA and HMI onboard the SDO, we have compared the EUV emissions, continuum intensity, emission measure, and temperature in the two kinds of regions. \\

We have seen that the EAPs in the moss area are associated with enhanced EUV emission measures in the logarithmic temperature range of 5.6-6.2.  We also find that  EUV emissions at 94, 131, 171, 193, 211, and 304 \AA\ over the EAPs are {\bf mostly} stronger than the corresponding emissions over the LAPs. However, there is no difference for 335 \AA\  emissions over the two regions. The observed emission strength of an EUV line depends on a combined effect of electron density, temperature, and response function of the AIA filters. The AIA channels at 94 \AA\ and 131 \AA\ are designed to study hot and extremely hot plasma in solar flares ($\sim$ 6 MK and 10 MK), yet the two lines do have additional low-temperature contributions \citep[e.g., ][]{2010A&A...521A..21O, 2011ApJ...732...81A}.  The AIA channel at 335 \AA\ is designed to study hot plasma in solar flares at 2.8 MK, and the response function of the 335 \AA\ filter in AIA is almost flat from $10^5$K to $10^7$K, being too wide to be useful for this analysis. \cite{2019ApJ...870..114S} reported that the photospheric continuum dominates the UV emissions observed by AIA in quiescent regions: 67\% in the AIA 1600 \AA\ filter and 87\% in the 1700 \AA\ filter. They also reported a broad range of temperature responses (4.2 $<$ log T $<$ 5.1) for the transmitted emissions from the two filters. The lack of distinction between the LAP and EAP regions in 1600 and 1700 \AA\ is likely due to the broader filters in these channels, sampling a wider temperature range. \\
 
To see why the mean temperature over the EAPs is lower than over the LAPs, the temperature of a certain region is estimated with differential emission measure weighted quantity as,

\begin{equation}
\bar{T}=\frac{\int_T DEM(T)TdT}{\int_T DEM(T) dT} \approx \frac{\Sigma DEM_{i,i+1}\frac{T_{i} + T_{i+1}}{2}}{\Sigma DEM_{i,i+1}} = \Sigma\frac{DEM_{i,i+1}}{\Sigma DEM_{i,i+1}}\frac{T_{i} + T_{i+1}}{2}
\end{equation}

 where $DEM_{i,i+1}$ is the differential emission measure in the temperature bin $(T_{i},T_{i+1})$.  We see that the lower temperature over the EAPs is due to the greater fraction of cool material there. The results may confirm an earlier finding that moss's electron density decreases in an active region with increasing temperature \citep{2008A&A...481L..53T}. In such a case, it's reasonable to think that PRM plays a dominant role in the orthohelium population in the moss region. The spatial structure of EAPs and LAPs may reflect the change in opacity caused by the chromospheric electron density, as we know from the results of a 3D radiation-MHD simulation by \cite[][and references therein]{2016A&A...594A.104L}. \\
 
Our results support the conclusion of the previous studies in that 10830 \AA\ EAPs are mostly located on intergranular lanes \citep{Ji2012ApJ, Hong2017RAA, Hashim2021RAA,  Ji2021RAA}. This is to say, over some parts of intergranular lanes, the material density of both cool and hot material is higher there. The EAPs are ubiquitous over solar active regions as well as quiet regions \citep{Hong2017RAA}. We still do not know what kind of activity causes the higher plasma density over there, but the process should be mild. The nature of ubiquitousness and mildness can account for the observed steadiness of corona heating as recently revealed by \cite{2023ApJ...957...25J}. The chromosphere is a highly dynamic layer, a sea of various jet phenomena to even smaller spatial scales than the current largest solar telescope can resolve, including spicules, fibrils,  mottles, and micro-jets, coupled with various MHD waves.  The EAPs may reflect small-scale dynamic activities like spicules in the chromosphere on some parts of intergranular lanes. Spicules are seen to channel hot plasma to the solar corona, and EUV emission generally appears at the top of the spicules \citep{2019Sci...366..890S, 2007PASJ...59S.655D, 2020SoPh..295..166Z}. Based on the above results, we suggest that more attention should be paid to some areas in intergranular lanes where heating channels, with enhanced density of both cool and hot material mingled together, may be located. Also, around the EAPs, the structure of plasma density and temperature is worth further investigation with more observations. \\

\section{Conclusion} \label{sec: Conclusion}

With high spatial resolution 10830 \AA\ filtergrams and nearly simultaneous data from multiple channels of the AIA and the continuum of the HMI on board SDO,
We analyzed an EUV moss area aiming to trace the clues of coronal heating channels down to the photosphere.  The filtergrams are obtained from the 1-meter aperture New Vacuum Solar Telescope (NVST), which is equipped with a near-infrared detector newly developed by the Kunming Institute of Physics and the Yunnan Astronomical Observatory.  We distinguish the darker regions from the brighter regions in the filtergrams and compare the EUV emissions, emission measure, temperature, and continuum intensity in the two kinds of regions. For convenience, we name the two kinds of regions in the moss area as enhanced absorption patches (EAPs) and low absorption patches (LAPs).  \\

 In short conclusion, over the EAPs for the EUV moss region analyzed in this paper, the mean EUV emission measure is higher than over the LAPs, but the mean temperature is lower. The lower temperature over the EAPs is due to the greater fraction of cool material there. Furthermore, over the EAPs, continuum intensity is lower, which supports the suggestion that the EAPs are located on intergranular lanes and might be the location of upward mass and energy into the corona. \\

%%%%%%%%%%%%%%%%%%%%%%%% Acknowledgments  %%%%%%%%%%%%%%%%%%%%%%%%%%%%%%%%%%
\section{Acknowledgments}
We thank the anonymous referee for helping us to improve the paper as well as SDO/AIA and SDO/HMI team for providing the high-quality data. The AIA and HMI data are downloaded via the VSO and the Joint Science Operations Center. This work is supported by the National Key R\&D Program of China 2021YFA1600502 (2021YFA1600500), the Natural Science Foundation (NSF) of Xinjiang Uyghur Autonomous Region (XUAR) (No. 2021D01A195), and West Light Foundation of CAS (No.2021-XBQNXZ-031).  This work was also supported by the Strategic Priority Research Program of the CAS Grant No. XDB0560000, and the NSF of China under grants 12273101 and 12003072. Thanks to the sponsorship by NSF of XUAR for support through the Outstanding Youth Project and “Tianchi Talents” special experts of XUAR.
%%%%%%%%%%%%%%%%%%%%%%%% References  %%%%%%%%%%%%%%%%%%%%%%%%%%%%%%%%%%

\nocite{*}
\bibliographystyle{aasjournal}
\bibliography{reference}

\end{CJK*}
\end{document}